# Bonding in Molecular Crystals from the Local Electronic Pressure Viewpoint

Vladimir G. Tsirelson [a], Adam I. Stash [b] and Ilya V. Tokatly[c,d,]

[a] *Quantum Chemistry Department, Mendeleev University of Chemical Technology, Moscow, 125047, Russia*
[b] *Laboratory of oxide materials, Karpov Research Institute of Physical Chemistry, Moscow, 103064, Russia*
[c] *Departamento de Fisica de Materials, Universidad del Pais Vasco UPV/EHU,20018 San Sebastian, Spain*
[d] *IKERBASQUE, Basgue Foundation for Science, E-20018, Bilbao, Spain*

Corresponding author: Vladimir G. Tsirelson. E-mail vtsirelson@yandex.ru, tel.+7 (499) 978-97-36

## Abstract

The spatial distribution of the internal pressure of an electron fluid, which spontaneously arises at the formation of a molecule or a crystal, is linked to the main features of chemical bonding in molecular crystals. The local pressure is approximately expressed in terms of the experimental electron density and its derivatives using the density functional formalism and is applied to identify the bonding features in benzene, formamide and chromium hexacarbonyl. We established how the spatial regions of compression and stretching of the electron fluid in these solids reflect the typical features of chemical bonds of different types. Thus, the internal electronic pressure can serve as a bonding descriptor, which has a clear physical meaning and reveals the specific features of variety of the chemical bonds expressing them in terms of the electron density.

## 1. Introduction

The hydrodynamic representation of one-particle Schrödinger equation is known since the early days of quantum mechanics [1, 2]. A many-body extension of this approach, which is known as "quantum fluid dynamics" (see [3] and references therein) or "quantum continuum mechanics" [4, 5], can be viewed as an orbital-free representation of the (current) density functional theory. From the continuum/fluid mechanics perspective the electrons in molecules and solids are considered as a charged elastic medium which moves and deforms under the action of external electromagnetic forces. In the equilibrium, the electronic continuum is stressed in such a way that the internal stress forces in the electronic subsystem balance the external electrostatic forces produced by the nuclei [6, 7, 8]. The corresponding electronic quantum-mechanical stress tensor carries all the necessary information about the internal spatial organization of this medium and therefore attracts much attention [9, 10, 11, 12, 13, 14, 15, 16, 17, 18, 19, 20, 21, 22, 23, 24, 25, 26, 27, 28, 29]. However,

the practical extraction of the chemically relevant information from this tensor is a nontrivial task for many-electron systems. Therefore, only a few examples for a limited number of systems are present in the literature [30, 31, 32, 33, 34, 35]. An additional complication is related to the well known nonuniqueness of the stress tensor [36, 27, 29].

Nevertheless, it is interesting to identify a physically relevant form of the electronic stress tensor and to establish a link between the quantum-mechanical stress in many-electron systems and the bonding models of classical structural chemistry. In particular, the main features of intra- and intermolecular bonds are determined by the spatial arrangement of the electrons in the outer (valence) electronic shells of atoms and molecules. The resulting picture of electron distribution follows from the electron-electron repulsion and electron-nuclear attraction [37]. Due to the Pauli principle, all electrons in closed-shell systems are proved to be distributed in pairs with opposite spins. The spatial arrangement of the electronic pairs and the degree of their localization, specific for each particular many-electron multi-nuclear system, can be interpreted in terms of regions of electron density (ED) concentration and depletion in the position space [38]. This concept is fruitful in the orbital models of structural chemistry as well [39, 40, 41].

However the real-space distribution of electron density does not show any indication on location of the electron pairs. Therefore, the different tools have been suggested to locate the regions of electron concentration and depletion. We mention the Laplacian of electron density [37], electron localization function (ELF) [42, 43], electron localizability indicator [44, 45, 46], localized orbital locator [47, 48], one-electron potential [49], maximum probability domains [50, 51, 52], conditional pair density [53], localized electron detector [54, 55, 56], single exponential decay detector [57, 58], information-theoretic ELF [59], steric [60] and Pauli [61] potentials and phase-space Fisher information density [62]. These descriptors play nowadays an important role in the chemical bonding analysis despite some their drawbacks. For example, the Laplacian of electron density does not display the uppermost electronic shells for many heavy atoms [63, 64] while LED and SEDD do not show the electron lone pairs. Therefore, the search for new physically motivated tools for extracting the bonding information from the electron density still remains in focus of many researches [65, 66].

To look at the situation from a new perspective [67, 68] we consider the picture of the internal quantum pressure of the inhomogeneous electron fluid. Thus function is simpler than the full electronic stress tensor; however it still contains significant bonding information. When a molecule or a crystal is formed, inhomogeneities of the electron density distribution are created in the space in a self-assembled manner. Accordingly, the internal stress and thus the pressure in the electronic continuum vary in the position space reflecting the spatial regions of compression and stretching. In this work we will establish a link between the distribution of the internal electronic pressure and the main chemical bond features in molecular crystals. Our motivation is twofold. First, the notion of the internal electronic pressure has a clear physical ground. Second, the distribution of pressure can be approximately expressed in terms of electron density and its derivatives: it allows us to use in computations the single-particle density in molecular crystals derived from accurate X-ray diffraction experiments [69]. The experimental density is N-representable, and the corresponding electronic internal pressure incorporates the Pauli principle requirements [6].

This paper is dedicated to Professor Andreas Savin on the occasion of his 65th birthday.

## 2. Methodology

Similarly to the classical continuum mechanics [70] , compression and stretching of quantum inhomogeneous electron fluid in molecules and crystals can be described in terms of a symmetric second-rank stress tensor, $\sigma_{ij}(\mathbf{r})$. Tensor's element $\sigma_{ij}$ corresponds to i-*th* component of the force acting on a unit surface perpendicular to j-axis. The divergence of this tensor

$$\frac{d\sigma_{ix}(\mathbf{r})}{dx} + \frac{d\sigma_{iy}(\mathbf{r})}{dy} + \frac{d\sigma_{iy}(\mathbf{r})}{dz}$$ defines i-*th* component of the force acting on an infinitesimal volume element of the electron fluid from outside. In the equilibrium state of a many-electron system, the internal stress forces determined by the divergence of $\sigma_{ij}(\mathbf{r})$ balance external forces.

Generically the electronic stress tensor $\sigma_{ij}(\mathbf{r})$ consists of the kinetic, exchange-corelation, and Hartree parts. As the Hartree contribution expected to be featureless [67] it is natural to extract it from the full stress tensor, and to consider the corresponding force as a part of the external one, together with the electrostatic force from the nuclei. The remaining part $p_{ij}(\mathbf{r})$ of the stress tensor describes the kinetic and the exchange-correlation effects. In the following tensor $p_{ij}(\mathbf{r})$ will be referred to as the internal electronic stress tensor. In the framework of the density-functional theory [71, 72], $p_{ij}(\mathbf{r})$ can be written as a sum of the Kohn-Sham kinetic and the exchange-correlation parts, $p_{ij}^{S}(\mathbf{r})$ and $p_{ij}^{xc}(\mathbf{r})$, respectively: $p_{ij}(\mathbf{r}) = p_{ij}^{S}(\mathbf{r}) + p_{ij}^{xc}(\mathbf{r})$ . Further, the internal stress tensor $p_{ij}(\mathbf{r})$ can be separated into the isotropic (bulk) part, $p(\mathbf{r})$, and the traceless (shear) part, $\pi_{ij}(\mathbf{r})$:

$$p_{ij}(\mathbf{r}) = \delta_{ij} p(\mathbf{r}) + \pi_{ij}(\mathbf{r}). \qquad (1)$$

By definition the function $p(\mathbf{r}) = \frac{1}{3} Tr\, p_{ij}(\mathbf{r})$ gives the **r**-distribution of the local average (isotropic) quantum pressure.

The general problem of working with the quantum stress tensor is the ambiguity of its definition [36, 27, 29]: any divergence-free tensor may be added to the stress tensor without changing the stress force. In practice a particular choice of the form for the internal stress tensor can be made based on physical arguments.

In this work we adopt the definition of the stress tensor which naturally follows from the formulation of the many-body theory in the co-moving Lagrangian frame [73, 74], and has been used recently to analyze the atomic shell structure [67]. The technical convenience of this definition is that it universally applies both to the kinetic and to the interaction parts of the stress tensor. In particular, the Kohn-Sham kinetic stress tensor reads [67]

$$p_{ij}^S(\mathbf{r}) = \frac{1}{2} \sum_{l\sigma} \left[ \frac{\partial \psi_{\sigma l}^*(\mathbf{r})}{\partial r_i} \frac{\partial \psi_{l\sigma}(\mathbf{r})}{\partial r_j} + \frac{\partial \psi_{\sigma l}^*(\mathbf{r})}{\partial r_j} \frac{\partial \psi_{l\sigma}(\mathbf{r})}{\partial r_i} \right] - \frac{1}{4} \delta_{ij} \nabla^2 \rho(\mathbf{r}), \quad i,j = x,y,z \quad (2)$$

where $\rho(\mathbf{r})$ is electron density, $\psi_{l\sigma}(\mathbf{r})$ are the Kohn-Sham spin-orbitas labeled by the index $l\sigma$; the atomic units are used. Accordingly, the kinetic part of the internal quantum hydrostatic pressure has the form

$$p^S(\mathbf{r}) = \frac{1}{3} \sum_{l\sigma} |\nabla \psi_{l\sigma}(\mathbf{r})|^2 - \frac{1}{4} \nabla^2 \rho(\mathbf{r}), \quad i,j = x,y,z. \quad (3)$$

Eq. (3) can be also represented as

$$p^S(\mathbf{r}) = \frac{2}{3} t_s(\mathbf{r}) - \frac{1}{4} \nabla^2 \rho(\mathbf{r}), \quad (4)$$

where $t_s(\mathbf{r})$ is density of kinetic energy of non-interacting Kohn-Sham particles.

If the exchange-correlation effects are described within GGA scheme, the corresponding exchange-correlation part of the quantum stress tensor, $p_{ij}^{xc}(\mathbf{r})$, takes the following form [67]:

$$p_{ij}^{xc}(\mathbf{r}) = \delta_{ij} [\rho(\mathbf{r}) v_{xc}(\mathbf{r}) - e_{xc}(\mathbf{r})] + \frac{\dfrac{\partial \rho(\mathbf{r})}{\partial r_i} \dfrac{\partial \rho(\mathbf{r})}{\partial r_j}}{2 k_F(\mathbf{r}) \rho(\mathbf{r}) |\nabla \rho(\mathbf{r})|} \frac{\partial e_{xc}(\mathbf{r})}{\partial s}. \quad (5)$$

Here $e_{xc}(\mathbf{r})$ is electronic exchange-correlation energy density defined according to

$E_{xc}[\rho] = \int e_{xc}(\mathbf{r}) d\mathbf{r}$; $s(\mathbf{r}) = |\nabla \rho(\mathbf{r})| / 2 k_F(\mathbf{r}) \rho(\mathbf{r})$ and $k_F(\mathbf{r}) = (3\pi^2 \rho(\mathbf{r}))^{1/3}$. Trace of the exchange-correlation stress tensor, $p_{ij}^{xc}(\mathbf{r})$, yields the internal pressure of inhomogeneous electron gas resulting from the exchange-correlation effects:

$$p^{xc}(\mathbf{r}) = \rho(\mathbf{r}) v_{xc}(\mathbf{r}) - e_{xc}(\mathbf{r}) + \frac{s}{3} \frac{\partial e_{xc}(\mathbf{r})}{\partial s}. \quad (6)$$

The main intuitive/interpretative advantage of the above definition of the local quantum stress tensor is that its physical content most closely resembles the physics behind the definition of macroscopic pressure in classical thermodynamics. The usual macroscopic pressure is defined as the response of the internal energy to changes of the volume at fixed number of particles. Similarly, the quantum stress tensor, Eq. (1), describes a change of the internal energy (virtual work) due to a virtual local deformation of an infinitesimal fluid element, provided the number of particles inside the deformable element is preserved. In particular, the quantum pressure $p(\mathbf{r})$ determines the virtual work under compression of a fluid element located at the point $\mathbf{r}$ while preserving its shape. Similarly, the shear part $\pi_{ij}(\mathbf{r})$ ($Tr\pi_{ij}(\mathbf{r}) = 0$) of the stress tensor describes the response to a local reshaping of the fluid element at $\mathbf{r}$ without changing its volume.

In the rest of this work we focus on the isotropic quantum pressure part of the internal electronic stress tensor. Further, we will take into account only exchange effects, assuming for simplicity that the electron correlation is weak. In the local density approximation (LDA) [71]

$$e^x_{LDA}(\mathbf{r}) = -\frac{3}{4}(\frac{3}{\pi})^{1/3}\rho^{1/3}(\mathbf{r}) \tag{7}$$

and LDA exchange contribution to the internal electron density pressure is

$$p^x_{LDA}(\mathbf{r}) = -\frac{1}{4}(\frac{3}{\pi})^{1/3}\rho^{4/3}(\mathbf{r}). \tag{8}$$

Thus, distribution of the isotropic internal pressure of the inhomogeneous electron fluid caused by quantum kinetic and exchange effects equals to $p(\mathbf{r}) = p^S(\mathbf{r}) + p^x(\mathbf{r})$, where $p^S(\mathbf{r})$ is defined by Eq. (4), while $p^x(\mathbf{r})$ is given in LDA by Eq. (8). Importantly, within the orbital-free DFT approach [75], these quantities can be expressed in terms of the ED, gradients of ED and the Laplacian of ED, and eventually derived from accurate X-ray diffraction experiments. For example, the function $t_s(\mathbf{r})$ can be approximated by the second-order gradient expansion [76] or by using some other similar DFT approximations, see [77]. Usually approximate functions $t_s(\mathbf{r})$ show a wrong asymptotic behavior close to the nuclei positions. Therefore small areas around the nuclei should be excluded from the consideration [78].

Thus, quantum internal pressure within the orbital-free DFT approach can be derived from accurate X-ray diffraction experiment. A detailed discussion of application of the experimental ED in the DFT formalism can be found in Refs. [77, 78, 79]. We used the experimental electron density and its derivatives expressed in terms of Hansen-Coppens space-extended multipole structural model [80]; in computations, the actual values of these quantities at each point $\mathbf{r}$ have been employed. Numerical multipole parameters for benzene, formamide and chromium hexacarbonyl are taken from [81], [82] and [83], correspondingly. The core and valence radial functions of the multipole model are approximated by the atomic wavefunctions listed in [84]. The function $t_s(\mathbf{r})$ was approximated according to [77]. All the calculations have been performed using the program WinXPRO [85, 86, 87], version 3.1.02.

### 3. Results and discussion

To qualitatively interpret the picture of internal electronic pressure in molecules and crystals, we note that positive or negative values of the pressure at the point $\mathbf{r}$ correspond, respectively, to the positive or negative virtual work under the local compression of an infinitesimal fluid element located at that spatial point. The corresponding regions of the electron fluid in a molecule or a crystal can be naturally interpreted as "compressed" (resistive to further compression) or "stretched" (tending to expand spontaneously).

In spherically symmetric free atoms the electronic shell structure is characterized by alternating positive and negative layers of pressure as noted in [67]. The former corresponds to the compressed electron fluid forming the electronic shells, while the latter is typical for the intershell (stretched) regions. The loss of spherical symmetry by the valence (outer) electronic shells of bounded atoms in molecules leads to the change in distribution of the electronic pressure with appearance of local maxima and minima in the space. The compressed regions correspond to the localized bonded and non-bonded electron pairs. Their number and mutual arrangement has to match the location of Lewis electron pairs in the VSEPR model of molecular geometry [39]. Similar behavior of electron pairs was early described in the literature in terms of Laplacian of electron density [40] and ELF [88].

We consider now the features of the internal pressure distribution of inhomogeneous electron fluid for different cases of bonding in a few molecular crystals.

### 3.1. Solid benzene

At normal conditions, benzene forms orthorhombic crystals (sp. gr. *Pbca*) with four differently oriented symmetry-independent molecules placed at the corners and at the centers of the three faces of the unit cell. Molecules are weakly bounded to each other; therefore we can consider the separate molecule "removed" from a crystal. Fig.1 shows the distribution of the internal electronic pressure that includes quantum kinetic and exchange effects, $p(\mathbf{r}) = p^S(\mathbf{r}) + p^x(\mathbf{r})$, in the plane of $C_6H_6$ molecule as well as in the cross-section perpendicular to this plane at the C–C bond critical point, the saddle point in electron density [37].

In benzene, the internal electronic pressure shows the alternation of positive and negative layers within atomic cores, reflecting the shell structure of bounded atoms. The saddle points in $p(\mathbf{r})$ are observed in C–C and C–H bond. The increase of compression of electron fluid, starting at the bond critical point of the covalent C–C interaction and growing towards C atoms, reflects the shift of the electrons to the atomic nuclei. Also, the compression spanning the bond axis towards the bond path (Fig.1,b) is seen. Remarkably, the pressure distribution, $p(\mathbf{r})$, shows the elongation of the compressed C–C middle-bond region perpendicular to the molecule plane of benzene (Fig.1, b). It is instructive to compare this picture with the distribution of the internal electron fluid pressure perpendicular to the ordinary C–C bond in diamond, see Fig. 1,b. It demonstrates that the σ-bond in diamond corresponds to an axially-symmetrical pressure distribution relatively the internuclear line accompanied with the electron compression towards that line. One naturally concludes that the aforementioned pressure pattern in benzene reflects the π-component of the C–C bond, a characteristic feature of benzene molecule in the standard orbital bonding models.

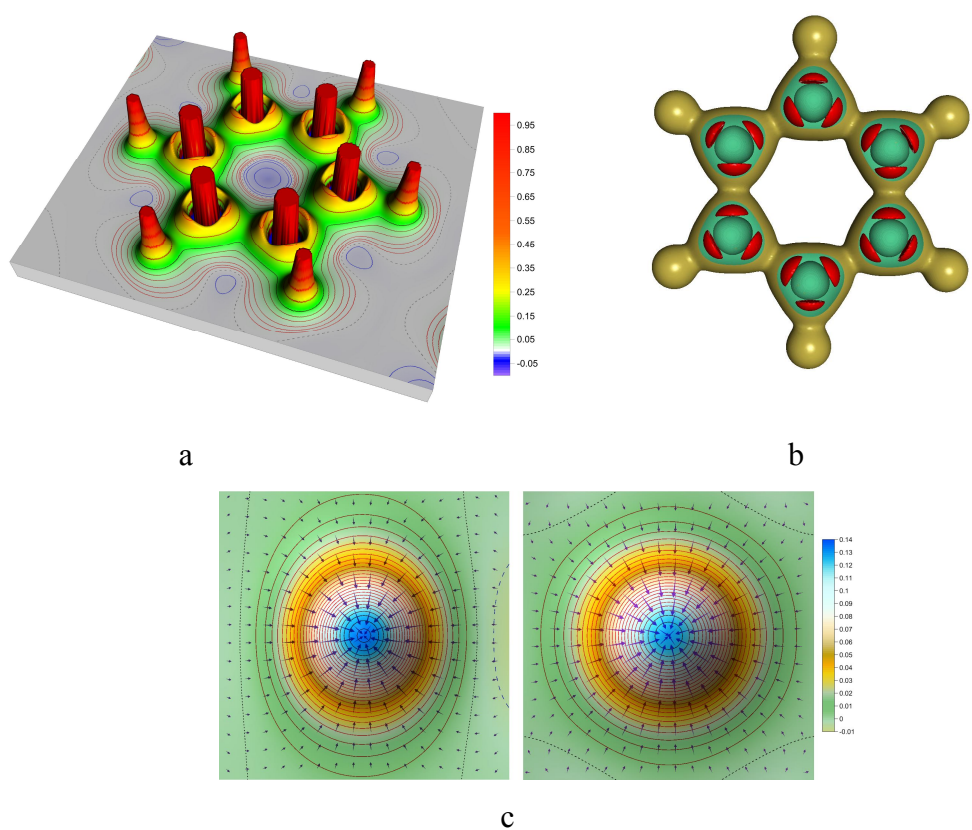

a

b

c

Figure 1. The distribution of electronic pressure in benzene $C_6H_6$. a) The molecular plane; b) the pressure distributions in cross-sections perpendicular to the C–C bonds in benzene (left) and diamond (right). Arrows show the pressure enhancement towards the bond axis; c) 3D picture showing the positive (compression) pressure regions of a molecule; the outer gold-yellow surface corresponds to the pressure p=0.1 a.u.; the red local maxima around C atoms cores reflecting the bonding electron concentrations are drawn at p=0.35 a.u.

The pressure distribution in the C–H bonds in benzene reveals the electron fluid compression along the bond line towards C atoms. This observation agrees with the empirical polarity of the C–H bond. Compression of electrons to C-H σ-bond axis takes place as well. At the same time, we notice an almost zero pressure region behind H atoms along to line of the C–H covalent bond. This feature favors participation of the H atom in intermolecular interactions in solid benzene.

The electronic pressure around the ring critical point at the centre of a benzene molecule is negative indicating stretching of the electron fluid, which is gradually changing to compression towards the atomic nuclei of the atomic ring. Also, we see the compression of electrons towards the molecule plane directed perpendicular to the plane of the molecule.

Thus, one can say that the benzene molecule skeleton is formed by the positive (compression) regions of a pressure. The enhanced local pressure regions are observed on the covalent bond lines: they reflect the bonding electron concentration (Fig. 1,c).

### 3.2. Formamide

The crystal structure of formamide, $HCONH_2$, consists of N–H...O hydrogen bonded zigzag chains [89], the chains are linked sideways by pairs of H-bonds connecting the formamide molecules. The couples of molecules form the cyclic dimers about a center of symmetry by means of weak N–H...O hydrogen bonds (2.947Å). The pressure distribution in the formamide dimer is shown in Fig. 2a. We see the electron fluid compression towards all the covalent bond and local maxima of enhanced pressure on the bond lines reflecting the bonding electron concentrations. The saddle points in $p(\mathbf{r})$ are observed in all the covalent bonds. At the C=O bond critical points, the pressure shows the compressed region elongated perpendicular to the molecule plane (figure is not shown); it agrees with the presence of π-component of the C=O. Such elongation of the compressed region for C–N bond is pronounced just very slightly, which is probably related to a small electron delocalization over the near-plane formamide molecule.

The local pressure maxima around all the atomic positions, reflecting the bonding electron concentrations, are seen in Fig. 2a. In addition, we notice the pressure maxima corresponding to nonbonding electron concentrations (electronic lone pairs) close to O atoms. Their mutual arrangement confirms the $sp^2$-hybridization of O atom.

It is worth noting that the above mentioned features (excluding the interatomic saddle points) are absent in the promolecular analog of the pressure in the fragment under discussion (Fig. 2b). The latter represents the pressure in the unbounded superposition of spherically symmetric atoms placed in real atomic positions (promolecule [69]). To make the bonding features in the pressure even more evident, we also computed the difference between actual pressure and that in the promolecule. The corresponding difference map (Fig.2c) reveals the enhancement of the actual pressure on the covalent bonds in a crystal with respects to promolecule and the reduction of pressure between the atoms which are not linked by the bond paths. It also shows the details of the nonbonding electron pair localization in the excessive pressure regions.

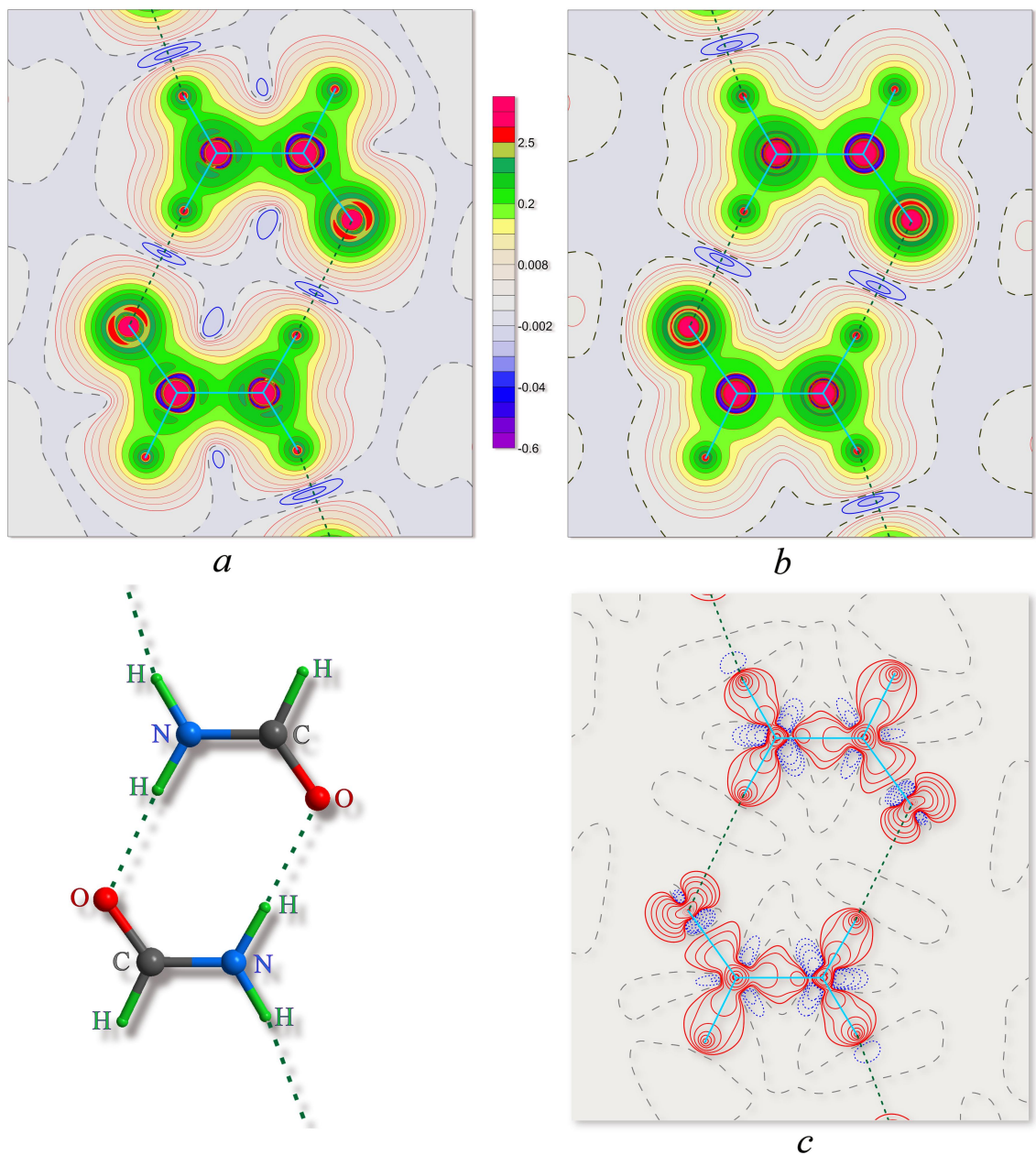

Figure 2. The electronic pressure distribution in formamide, $HCONH_2$; the bounded molecular dimer in a crystal is shown: the black dashed lines indicate the weak hydrogen bonds.  a) the pressure distribution in the crystal fragment; b) pressure distribution in the promolecular analog of this fragment; c) the difference between actual and promolecular electronic pressure; the red lines indicate the excessive pressure regions, the line intervals are $\pm 2 \times 10^n$, $\pm 4 \times 10^n$ and $\pm 8 \times 10^n$ a.u. (n= -2, -1).

An interesting feature of Fig. 2a is the negative local pressure areas on the weak N–H...O hydrogen bonds linking the molecules in the dimers. They are observed in the promolecular map as well. The change in the pressure on these N–H...O hydrogen bonds at crystal formation from atoms is very small. Moreover, the function $p(\mathbf{r})$ does not show the saddle point in the weak hydrogen bonds, in contrast to the picture found for $p(\mathbf{r})$ on all the covalent bonds above. We withhold comments at this stage. The point is that the pressure is an averaged quantity as the shear part is traced out of the full stress tensor. However, the shear deformation may be significant for hydrogen bonds, for which pressure is negative and a large pressure enhancement from the bond critical points towards atoms does not take place. The full stress tensor is needed to extract the exhaustive bonding information and it would be interesting to learn which features of the stress distribution will be obtained in this case. We plan to address this point in the nearest future.

### 3.3. Cr(CO)6

Let us now consider the electron pressure features reflecting the metal-ligand interactions in chromium hexacarbonyl, $Cr(CO)_6$. Single Cr atom exhibits the three pairs of positive and negative electron pressure areas, the outermost positive area is in the range 0.27 <r< 0.66 Å. In a molecule, the spherical symmetry of the atom is destroyed because of the bonding effects.  Fig. 3 demonstrates that the eight enhanced pressure regions are formed in $Cr(CO)_6$ in the corners of the cube surrounding the Cr atom. At the same time, the low pressure or stretched areas ("the holes") appear in the faces of this cube.  The nonbonded compression regions behind the C atoms in CO ligands show the mutual alignment with these holes. This picture fits the "*key and lock*" metal-ligand σ-bonding mechanism in $Cr(CO)_6$ [90].

The described results agree excellently with those obtained from the Laplacian of both theoretical [90] and experimental [83] electron density of $Cr(CO)_6$. The regions of enhanced pressure in the corners of the cube surrounding the Cr atom coincide with the ED concentrations as revealed by the Laplacian. Also, the location of the maxima and minima in the electronic pressure distribution around a transition metal atom in $Cr(CO)_6$ agrees with the transfer of ED from Cr to the π*-orbitals of the ligands as it follows from the Chatt-Duncanson $d_\pi$-$p_\pi$* back-donation model [91]. Indeed, according to the crystal field theory [92], the three $t_{2g}$ atomic orbitals (AOs) occupied by electrons in a chromium complex are directed along the cube diagonals, while the two empty $e_g$-AOs show up towards the six faces of this cube. By the symmetry, the $e_g$-AOs of Cr atom interfere with σ-orbitals of ligands while $t_{2g}$-AO of metal intercross the ligand π-orbitals. The eight electron fluid compression regions along the cube diagonals in the pressure distribution around Cr atom can be associated with the electron-populated $t_{2g}$-orbitals that interact with the π*-orbitals of the ligands.

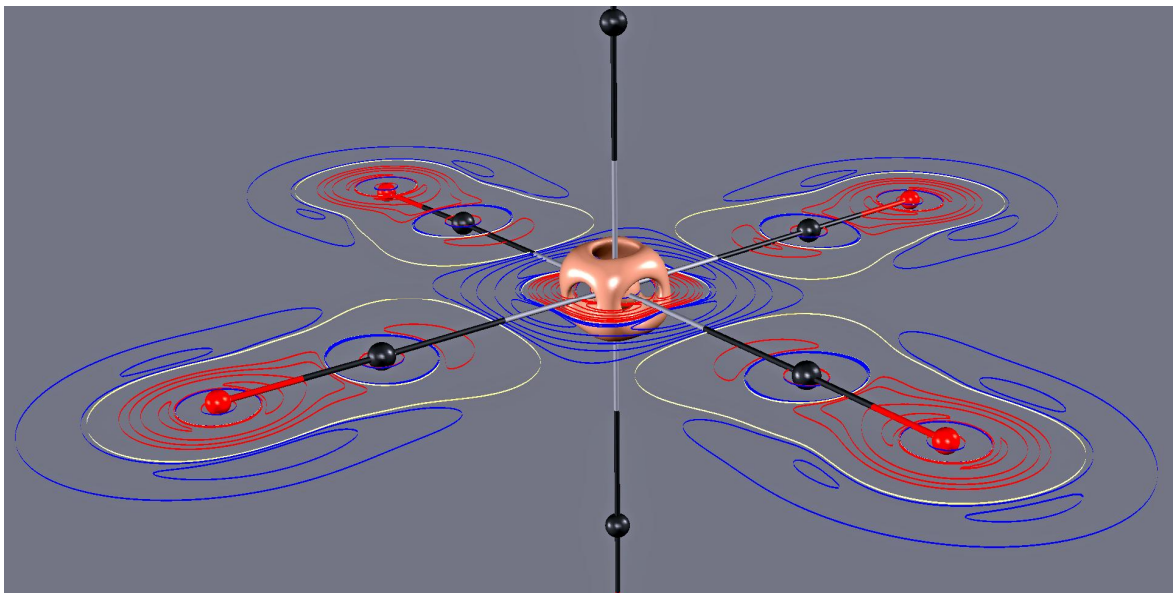

Figure 3. The electronic pressure distribution in the Cr(CO)$_6$ molecule extracted from a crystal. The positive and negative pressure regions are marked by the red and blue lines, respectively. The pressure in the cubic-shaped region around the Cr atom at the center of the molecule corresponds to p=8 a.u.

Thus, the features of the internal electronic pressure give the orbital-free pattern in agreement with that predicted by the simple crystal field theory [92]. They reveal a donor–acceptor interaction details in terms of physical orbital-free quantities.

Finally, we notice the pressure enhancement in the triple C≡O bond, where the saddle point is observed, and a nonbonding pressure maximum (the electronic lone pair) behind O atom along the C–O line.

## 4. Summary

The internal pressure of the inhomogeneous electron fluid in molecules and crystals can be viewed as the function describing the system response to a virtual isotropic compression of an infinitesimal electronic fluid element located at the point **r**. The sign of the pressure indicates what exactly – compression or stretching – takes place. These compressed/stretched areas correspond to the electron density concentration and depletion regions as revealed by the Laplacian of the electron density or ELF. They are also in agreement with bonding information from the orbital chemical bond models. Thus, the typical features of chemical bonds of different types are reflected in the electronic pressure distribution. For example, the homoatomic covalent C–C bonds show the compression of electrons towards the bond axes, the bonds of formal bond order of 1.0 (diamond) and 1.5 (benzene) exhibit specific features consistent with those in the orbital model of bonding.

The bonding and lone electron pairs manifest themselves as the local areas of enhanced electronic pressure. While all these features are not explicitly visible in the electron density itself, they become evident in terms of the electronic pressure. Thus, this function can serve as a bonding descriptor, which decodes the chemical bonding information hidden in the electron density. It is important that, in contrast to many other bonding descriptors, the internal electronic pressure has a clear physical meaning and it is linked directly to the classical consideration of deformable media. Moreover, it is derivable from the experimental electron density and its derivatives using transparent orbital-free DFT approximations. It is worth emphasizing that both the kinetic and exchange-correlation contributions to the pressure can be separately obtained in the simple and natural way. Finally, our approach provides a simple and natural way to study the bonding features in crystals under external pressure [93, 94, 95, 96, 97, 98, 99].

We presented here the local electronic pressure analysis of bonding. The part of the bonding information related to the shear deformations does not appear in this scheme; at the same time it may be significant for the van der Waals interactions and weak hydrogen bonds, for which a large pressure shift towards the atoms from the bond critical points is observed. The total stress tensor, $p_{ij}(\mathbf{r})$, is needed to elucidate this matter. It is important to indicate that development of our approach will allow to reconstruct $p_{ij}(\mathbf{r})$, i.e., to determine the eigenvalues and the principal axes of this tensor. By visualizing its topology and analyzing its characteristics we can expect to get even more information about chemical bonding in molecules and crystals. In particular, our form of the stress tensor, Eqs. (2) and (5), which differs from that used in the works [32, 100, 101], comes from a physically reasonable treatment of the internal electronic stress in terms of the virtual work under a local deformation of infinitesimal fluid elements. It is natural to expect that anisotropy of the response to various local deformations, which is encoded in the tensor structure, should reflect the anisotropy of chemical bond both along and perpendicular to the bond path. Therefore, it would be interesting to apply our approach to the problem of the formation and breaking of bond paths in molecules and solids. This work is now in progress in our group.

**Acknowledgement**

We thank Dr. D. Yufit, Prof. S. Parsons and Dr. L. Farrugia for sending us the experimental data for benzene, formamide and chromium hexacarbonyl. This work was supported by the Russian Foundation for Basic Research under grant 13-03-00767a (VGT and AIS); Spanish Ministerio de Economía y Competitividad (MINECO) under grant No. FIS2013-46159-C3-1-P and the "Grupos Consolidados UPV/EHU del Gobierno Vasco" under Grant No. IT578-13 (I.V.T).